\documentstyle [twocolumn,aps,prb,epsf,psfig]{revtex}
\begin{document}
\title{Shadow bands, gap and pseudogaps in high-$T_c$ superconductors.}

\author{S. Caprara, A. Perali, M. Sulpizi}

\address{Dipartimento di Fisica, Universit\`a di Roma ``La Sapienza'', 
and INFM Unit\`a di Roma 1,
P.le A. Moro 2, 00185 Roma, Italy}
\maketitle
\begin{abstract}
Within the framework of the Charge Density Wave Quantum Critical Point 
($CDW$-$QCP$) scenario for high-$T_c$ superconductors (HTCS),
we introduce a model for tight-binding electrons coupled
to quasi-critical fluctuations.
In the normal state our model reproduces 
features of the Fermi Surface (FS) observed in $ARPES$ 
measurements on optimally doped Bi2212, such as
the anisotropic suppression of spectral weight 
around the $M$ points of the Brillouin zone. The
spectral density is characterized by a transfer of spectral weight
fro the main quasi-particle peak
to dispersing shadow peaks which originate branches of a shadow FS.
  In the superconducting state our model reproduces the 
$d$-wave symmetry of the gap parameter, which
results from a balance between small-$q$ attraction and large-$q$
repulsion. The gap parameter is enhanced due to cooperative effects
of charge and spin fluctuations.
\end{abstract}

\vspace{1cm}

Recently it has been proposed that the anomalous normal-state
and the superconducting properties of HTSC
are the natural consequence of the existence
of a $CDW$-$QCP$ near optimal doping [1-4] which gives rise
to large charge fluctuations in the electronic system. 
The dynamical charge segregation in hole-poor and hole-rich regions
induces antiferromagnetic spin fluctuations even at optimal doping.
When the electrons are coupled to quasi-critical charge and spin fluctuations
a strongly momentum and frequency dependent electron-electron interaction 
appears.

We consider a model of tight binding electrons coupled with
charge and spin fluctuations
$$
H=\sum_{k,\sigma}\xi_k c^{\dagger}_{k,\sigma}c_{k,\sigma}
+\sum_{j=0}^4 g_j\sum_{k,q,\sigma,\sigma'}
c^{\dagger}_{k,\sigma}c_{k+q,\sigma'}\tau^j_{\sigma,\sigma'}S^j_{-q}
$$
where $\xi_k=-2t(\cos k_x+\cos k_y+\alpha \cos k_x \cos k_y)-\mu$
(commonly used values for Bi2212 being $t=0.2 eV$, $\alpha=-0.5$
and $\mu=-0.18 eV$ to fix the
hole doping $\delta=0.17$), 
$j=0$ for charge and $j=1,2,3$ for spin.
  
The fluctuating modes are characterized by effective susceptibilities
$$
\chi_{ij}(q,\omega)=\delta_{ij}\frac{A_j}{\Omega_j(q)-{\rm i}\omega }
$$
where $\Omega_j(q)=t^*_j+\bar{\omega}_j\left[2-\cos(q-Q_j)_x-\cos(q-Q_j)_y
\right]$, $t^*_j$ is a mass term depending on the distance from the $QCP$,
$\bar{\omega}_j$ is a characteristic energy scale associated with
the damping of the fluctuations, and the critical wavevector
for spin fluctuations is $Q_s\simeq(\pi;\pi)$ while for 
charge fluctuations we take
$Q_c=(0.4\pi;-0.4\pi)$ as suggested by recent experimental results [5]. 
The above expression 
for the effective spin susceptibility was suggested to fit NMR
results in the region of overdamped spin waves [6]. The form of
the charge susceptibility results from a slave-boson calculation within the
Hubbard-Holstein model with long-range Coulombic interaction, close
to the CDW instability [1].

We calculate first order selfenergy corrections to the quasiparticle spectra.
The imaginary part of the selfenergy is characterized by peaks corresponding
to the energies $\epsilon\simeq\xi_{k-Q_j}$; 
as $\xi_{k-Q_j}\rightarrow 0$ the peaks 
are suppressed while the usual low-energy Fermi-liquid behavior is turned into 
an anomalous square-root behavior.
The peaks in the imaginary part of the 
selfenergy are responsible for the appearance
of shadow peaks in the spectral density
$$
A(k,\omega)=\frac{1}{\pi}\frac{|Im\Sigma(k,\omega)|}
{\left[\omega-\xi_k-Re\Sigma(k,\omega)\right]^2
+\left[Im\Sigma(k,\omega)\right]^2} 
$$

To compare with the $ARPES$ experiments in [5] we describe the FS
associating to each $k$-point the spectral 
weight $p_k$ of occupied states within an energy window $[-W,W]$
(a typical experimental value is $2W=50 meV$), i.e.

$$
p_k=\int_{-W}^{W}d\omega A(k,\omega)f(\omega).
$$

Spin fluctuations by themselves would preserve the full symmetry of the 
FS due to the peculiar commensurability of the critical wavevector
$Q_s=(\pi;\pi)$. 
The main effects of the coupling to spin fluctuations is a 
symmetric suppression 
of spectral weight around the $M$ points of the Brillouin zone, and 
the appearance of weak ``hole pockets" around the points $(\pm\pi/2;\pm\pi/2)$
due to a transfer of spectral weight to lower energies at the shadow FS.
Along the $\Gamma M$ direction the main quasiparticle peak is suppressed and a 
broader shadow peak appears at the M point below the Fermi energy.

In Bi2212 the experimentally observed FS [2] indicates that the 
mirror symmetry 
with respect the $\Gamma X(Y)$ axes is preserved, whereas the symmetry 
with respect to the $\Gamma M(M_1)$ axes is broken, suggesting a $Q_c$ 
directed  along the $\Gamma X(Y)$ directions. A charge-fluctuation 
mode with a diagonal $Q_c$ may reproduce these features,
the main effect of charge fluctuations being, indeed,
an asymmetric modulation of 
spectral weight around the $M(M_1)$ points.
When both modes are taken into account, 
the suppression of spectral weight around the $M$ points 
due to spin fluctuations is modulated 
by charge fluctuations leading to an asymmetric distribution as 
experimentally observed (Fig. 1).
Moreover the broad structure at the $M$ point below the Fermi energy,
due to spin fluctuations, persists.

We then consider the static effective interaction 
in the Cooper channel  
$\Gamma_{eff}(q)=g_s^2\chi _s (q,\omega =0)-g_c^2\chi _c (q,\omega =0)$, 
and we solve the $BCS$ equation
$$
\Delta(k)=-\frac{1}{N}\sum_{p}\Gamma_{eff}(k-p)
\frac{\tanh\frac{\varepsilon_{p}}{2T}}{2\varepsilon_{p}}
\Delta(p)
$$
where $\varepsilon_{p}^2=\xi^2_{p}+\Delta(p)^2$ and
$\Delta(k)$ is the gap parameter.
The cooperative charge and spin fluctuations enhance the 
$d_{x^2-y^2}$-wave gap parameter leading to larger values 
with respect to the
charge or spin fluctuations when considered separately.
The $k$ dependence of the gap is due to an interplay between
the band structure and the effective interaction. The modulus of the gap tends
to follow the local density of states for small $q$ attraction.
For singular quasiparticle interactions the gap has local maxima
at the points where $\xi_k=\xi_{k-Q_j}=0$
(hot spots). We obtain two different behaviors for $\Delta (k)$
in the $d_{x^2-y^2}$ channel according to vertical or diagonal  $Q_c$ 
directions (Fig. 2).

In the underdoped region,the charge fluctuation become critical near the 
$T_{CDW}$ line, where $CDW$ transition would occur in the absence of pairing
and the attractive fluctuations lead to local pairs with pseudogap
opening. The stabilizing effect of local superconducting order, with respect
to the $CDW$ instability, is introduced via the local gap. In this 
situation we reproduce the general trend of the temperature dependence
of the pseudogap.     

In {\bf conclusion},
 within the framework of the $CDW$-$QCP$ scenario for HTSC,
we have investigated the properties of a model for electrons coupled
with both charge and spin quasi-critical fluctuations.

In the normal state we reproduce features of the FS as 
observed in recent $ARPES$ measurements [5,7].
In particular we recover an anisotropic suppression of spectral weight 
around the $M$ points of the Brillouin zone and shadow features 
associated to charge and spin fluctuations.

In the superconducting state we obtain a $d$-wave superconducting gap
as a result of the balance between repulsion (spin fluctuations) and 
attraction (charge fluctuations) between
quasiparticles. We find that
the gap parameter and the critical temperature are enhanced due to
the cooperative effects of charge and spin fluctuations.\\
\par\noindent
{\bf Acknowledgements:} The results presented in this paper 
are the highlights of a
work done in collaboration with C. Di Castro, C. Castellani and M. Grilli.
Useful discussions with A. Bianconi are acknowledged. Two of us (S.C. and M.S.)
are supported by the INFM - PRA 1996.\\

\par\noindent
\par\noindent
{\bf References}
\par\noindent
 $[1]$ C. Castellani {\em et al.}, Phys. Rev. Lett. {\bf 75}, 4650 (1995).\\
 $[2]$ A. Perali {\em et al.}, Phys. Rev. B {\bf 54}, 16216 (1996).\\
 $[3]$ C. Castellani {\em et al.}, Z. Phys. B {\bf 107}, 137 (1997).\\
 $[4]$ S. Caprara {\em et al.}, 
 submitted to Phys. Rev. B (1998) cond-mat 9811130.\\
 $[5]$ N. L. Saini {\em et al.}, Phys. Rev. Lett. {\bf 79}, 3467 (1997).\\
 $[6]$ A. J. Millis {\em et al.}, Phys. Rev. B {\bf 42}, 167 (1990).\\
 $[7]$ D. S. Marshall {\em et al.},Phys. Rev. Lett. {\bf 76}, 4841 (1996).\\
 $[8]$ H. Ding {\em et al}, Nature {\bf 382}, 51 (1996).

%___________________________________________________________________________

\begin{figure}[htbp]   % produce figure here(top,bottom,page_float)
    \begin{center}
       \setlength{\unitlength}{1truecm}
       \begin{picture}(5.0,15.0)
          \put(-8.0,0.0){\epsfbox{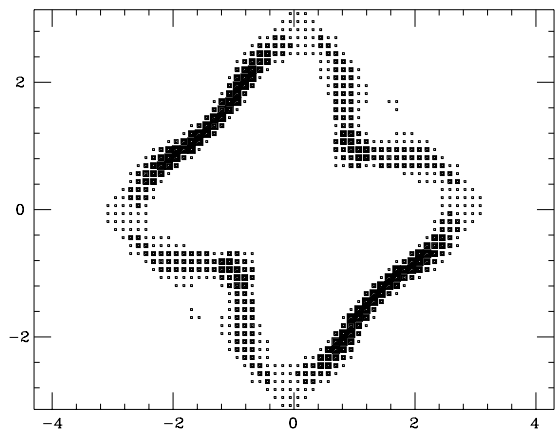}}    
       \end{picture}
    \end{center}
    \caption{Fermi surface in presence of charge and spin fluctuations
             with $Qc=(0.4\pi; -0.4\pi)$ and $Q_s=(\pi; \pi)$.}  
    \protect\label{fs}
\end{figure}

\begin{figure}
\protect
\centerline{\psfig{figure=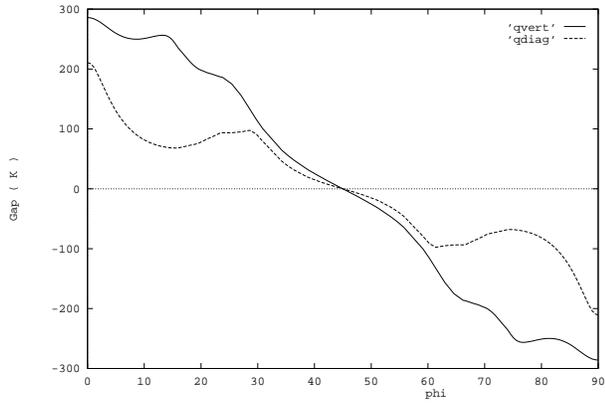,width=8.2cm,angle=-90}}
\caption{Gap parameter in presence of charge and spin 
fluctuations with $Q_c=(0.4\pi; 0)$ (solid line) or 
$Q_c=(0.4\pi; -0.4\pi)$ (dashed line) and $Q_s=(\pi,\pi)$.}
\end{figure} 

\end{document}